\documentstyle[12pt]{article}

\begin{document}

\begin{center}

\bigskip
{\Large Entanglement of Formation of an
Arbitrary State of Two Qubits}\\

\bigskip

William K.~Wootters\\

\bigskip

{\small{\sl

Department of Physics, Williams College,
Williamstown MA 01267}}  \vspace{2mm}

\end{center}
\subsection*{\centering Abstract}
{The entanglement of a pure state of a pair of quantum
systems is defined
as the entropy of either member of the pair.
The entanglement of formation
of a mixed state $\rho$ is defined as the minimum average
entanglement of an ensemble of pure states that
represents $\rho$.  An earlier paper
[{\em Phys. Rev. Lett.} {\bf 78}, 5022 (1997)] conjectured
an explicit formula for the entanglement of formation of a pair of
{\em binary} quantum objects (qubits)
as a function of their density matrix, and proved
the formula to be true for
a special class of mixed states.  The present paper extends the
proof to arbitrary states of this system
and shows how to construct entanglement-minimizing
pure-state decompositions.}

\vfill

{\noindent}PACS numbers: 03.65.Bz, 89.70.+c\vfill

\newpage
Entanglement is the feature of quantum mechanics that allows,
in principle, feats such as teleportation \cite{teleport}
and dense coding \cite{dense} and is what
Schr\"odinger called ``{\em the} characteristic trait
of quantum mechanics \cite{Schrod}.''
A pure state of a pair of quantum systems is called
entangled if it does not factorize, that is,
if each separate system does not have a pure state
of its own.  A classic example is the singlet state
of two spin-$\frac{1}{2}$ particles,
$\frac{1}{\sqrt{2}}(|\uparrow \downarrow \rangle -
|\downarrow \uparrow \rangle )$, in which neither particle
has a definite spin direction.
A {\em mixed} state is entangled if it cannot
be represented as a mixture of factorizable pure states.
In the last couple of years a good deal of work has been devoted to
finding physically motivated measures of entanglement, particularly
for mixed states of a bipartite
system \cite{pure,measures,formation}.
Perhaps the
most basic of these measures is the {\em entanglement of formation},
which is intended to quantify the resources needed to
create a given entangled state \cite{formation}.

Having a well justified and mathematically tractable measure
of entanglement
is likely to be of value in a number of areas of research,
including the study
of decoherence in quantum computers \cite{computers}
and the evaluation of quantum
cryptographic schemes \cite{crypto}.  Unfortunately, most proposed
measures of
entanglement involve extremizations which
are difficult to handle analytically; the entanglement of formation
is no exception to this rule.  However, in the special case of
entanglement between two {\em binary} quantum systems such as
the spin of a spin-$\frac{1}{2}$ particle or the polarization of
a photon---systems that are generically called
``qubits''---an explicit formula
for the entanglement of formation has recently been conjectured and has
been proved for a special class of density
matrices \cite{HW}.  In this Letter we
prove the formula for arbitrary states of two qubits.

The entanglement of formation is defined as follows
\cite{formation}.  Given a density
matrix $\rho$ of a pair of quantum systems $A$ and $B$, consider all
possible pure-state decompositions of $\rho$, that is, all ensembles
of states $|\psi_i \rangle$ with probabilities $p_i$ such that

\begin{equation}
\rho = \sum_i p_i |\psi_i \rangle \langle \psi_i| .
\end{equation}
For each pure state, the entanglement $E$ is defined as the entropy of
either of the two subsystems $A$ and $B$ \cite{pure}:

\begin{equation}
E(\psi) = - {\mbox{Tr}\,} (\rho_A \log_2 \rho_A) =
- {\mbox{Tr}\,} (\rho_B \log_2 \rho_B) .  \label{pure}
\end{equation}
Here $\rho_A$ is the partial trace of $|\psi\rangle\langle\psi|$ over
subsystem $B$, and $\rho_B$ is defined similarly.
The entanglement of formation of the mixed state $\rho$ is
then defined as the
average entanglement of the pure states of the decomposition,
minimized over all decompositions of $\rho$:

\begin{equation}
E(\rho) = \mbox{min}\, \sum_i p_i E(\psi_i) .  \label{mixed}
\end{equation}
The basic equation (\ref{pure}) is justified by
the physical interconvertibility of a collection of pairs
in an arbitrary pure
state $|\psi\rangle$ and a collection of pairs in the
standard singlet state,
the asymptotic conversion ratio being given by $E(\psi)$ \cite{pure}.
The central claim of this Letter is that for a pair of qubits,
the minimum value
specified in Eq.~(\ref{mixed}) can be
expressed as an explicit function of $\rho$, which we develop in
the next few paragraphs.  For ease of expression we will
usually refer to the entanglement of formation
simply as ``the entanglement.''

Our formula for entanglement makes use of what can be called
the ``spin flip'' transformation, which is a function applicable both
to state vectors and to density matrices of an arbitrary number of
qubits.  For a pure state of a single qubit, the spin flip, which we
denote by a tilde, is defined by

\begin{equation}
|\tilde{\psi}\rangle = \sigma_y |\psi^* \rangle ,
\end{equation}
where $|\psi^* \rangle$ is the complex conjugate of $|\psi \rangle$
when it is expressed in a fixed basis such as
$\{|\uparrow\rangle, |\downarrow\rangle \}$, and $\sigma_y$ expressed
in that same basis is the matrix $\pmatrix{0 &-i \cr i &0 \cr}$.
For a spin-$\frac{1}{2}$ particle this is the standard time reversal
operation and indeed reverses
the direction of the spin \cite{Sakurai}.
To perform a spin flip on
$n$ qubits, one
applies the above transformation to each individual qubit.
If the system is described by a density matrix rather
than a state vector,
each $\sigma_y$ is applied on both the right and the left.
For example, for a general state $\rho$ of two qubits---the
object of interest in this Letter---the spin-flipped state is

\begin{equation}
\tilde{\rho} = (\sigma_2 \otimes \sigma_2) \rho^*
     (\sigma_2 \otimes \sigma_2),
\end{equation}
where again the complex conjugate is taken in the standard basis,
which for a pair of spin-$\frac{1}{2}$ particles is
$\{ |\uparrow \uparrow \rangle, |\uparrow \downarrow \rangle,
|\downarrow \uparrow \rangle, |\downarrow \downarrow \rangle \}$.
In this case the spin flip
is equivalent \cite{CK} to ``complex conjugation
in the magic basis,'' which appears in Ref.~\cite{HW}.

Though we have introduced the spin flip transformation
primarily to deal with
mixed states, the concept is also convenient
for expressing the entanglement
of a {\em pure} state of two qubits.  One can show that
this entanglement,
defined in Eq.~(\ref{pure}), can be written as \cite{HW}

\begin{equation}
E(\psi) = {\cal E}(C(\psi)),  \label{E(psi)}
\end{equation}
where the ``concurrence'' $C$ is defined as

\begin{equation}
C(\psi) = |\langle \psi | \tilde{\psi} \rangle |, \label{C(psi)}
\end{equation}
and the function ${\cal E}$ is given by

\begin{equation}
{\cal E}(C) = - \frac{1+\sqrt{1-C^2}}{2} \log_2 \frac{1+\sqrt{1-C^2}}{2}
              - \frac{1-\sqrt{1-C^2}}{2} \log_2 \frac{1-\sqrt{1-C^2}}{2}.
\label{EofC}
\end{equation}
The function ${\cal E}(C)$ is monotonically increasing, and
ranges from 0 to 1 as $C$ goes from 0 to 1, so that one can take the
concurrence as a measure of entanglement in its own right.  For example,
the singlet state
$|~\psi~\rangle~=$
$\frac{1}{\sqrt{2}}(|\uparrow\downarrow\rangle-|\downarrow\uparrow\rangle)$
is left unchanged by a
spin flip (except for an overall negative sign), so that its concurrence
$|\langle \psi | \tilde{\psi} \rangle |$ is equal to 1.
At the other extreme, an unentangled, or factorizable,
pure state such as $|\uparrow \downarrow \rangle$ is always mapped by
the spin flip transformation into an orthogonal state, so that its
concurrence is zero.  Later we
will use another fact about ${\cal E}(C)$, namely,
that it is a convex
function (that is, curving upward).

Having defined the spin flip and the function ${\cal E}(C)$, we can now
present the promised formula for the entanglement of formation
of a mixed state $\rho$ of two qubits:

\begin{equation}
E(\rho) = {\cal E}(C(\rho)),  \label{E(rho)}
\end{equation}
where

\begin{equation}
C(\rho) = \mbox{max}\,\{0, \lambda_1-\lambda_2-\lambda_3-\lambda_4 \},
\label{C(rho)}
\end{equation}
and the $\lambda_i$s are the eigenvalues, in decreasing order, of the
Hermitian
matrix $R \equiv \sqrt{\sqrt{\rho}\tilde{\rho}\sqrt{\rho}}$.
Alternatively, one
can say that the $\lambda_i$s are the square roots of the eigenvalues
of the non-Hermitian matrix $\rho \tilde{\rho}$.  Note that
each $\lambda_i$
is a non-negative real number.  The matrices $R$
and $\rho \tilde{\rho}$ may seem unlikely objects to be using
in any formula, but
one can see that they are closely related to the pure-state concurrence
of Eq.~(\ref{C(psi)}).  In fact for a pure state $|\psi \rangle$,
$R$ has only one eigenvalue that may be nonzero, namely, $C(\psi)$.

The formula (\ref{E(rho)})
was shown in Ref.~\cite{HW} to be correct for all
density matrices
of two qubits having no more than two nonzero eigenvalues.
More recently, Smolin has tested the formula
numerically on several thousand randomly chosen
two-qubit density matrices
and has found complete agreement \cite{Smolin}.
Most of the rest of this Letter is devoted to proving that the
formula is correct for arbitrary states of this system.
We will find that the
value ${\cal E}(C(\rho))$ of the average entanglement
can always be achieved
by a decomposition of $\rho$ consisting of four or fewer pure states,
each state having the same entanglement.
(Uhlmann has already shown that the optimal
decomposition must consist of pure states with equal entanglement
\cite{Uhlmann}, but we do not assume this result in our proof.)  We
will then show that no decomposition has a smaller
average entanglement.

Our method will be to look explicitly for
an entanglement-minimizing decomposition of $\rho$.  We use the fact
that {\em every} decomposition
of a density matrix can be obtained via the following
prescription \cite{HJW}.  First,
find a complete set of orthogonal eigenvectors $|v_i\rangle$
corresponding to the nonzero eigenvalues of $\rho$, and
``subnormalize'' these vectors so that
$\langle v_i|v_i \rangle$ is equal to the $i$th eigenvalue.
Every decomposition $\{ |w_i\rangle \}$ of $\rho$
can then be obtained through the following equation, and every set of
states that can be obtained in this
way is a legitimate decomposition of $\rho$:

\begin{equation}
| w_i \rangle = \sum_{j=1}^n U_{ij}^* | v_j\rangle.  \label{w}
\end{equation}
Here $n$ is the rank of $\rho$, that is, the number of vectors
$|v_i\rangle$, and $U$ is an $m \times m$
unitary matrix, $m$
being greater than or equal to $n$.  (The complex conjugation is
included only for later convenience.)  Alternatively, since
only the first $n$ columns of $U$ are used, we can take $U$
to be an $m \times n$ matrix whose columns are orthonormal vectors.
If $m$ is greater than $n$,
the decomposition will have more
elements than are necessary for the creation of $\rho$, but such
decompositions are certainly allowed.  The states $|w_i\rangle$ in
Eq.~(\ref{w}) are automatically
subnormalized so that $\langle w_i|w_i \rangle$
is equal to the probability of $|w_i\rangle$ in the decomposition.
We can thus write $\rho = \sum_i |w_i\rangle \langle w_i|$.  In what
follows, we express all decompositions of $\rho$ in terms of
such subnormalized vectors.

It is helpful to consider separately two classes of density
matrix: (i) those for which $\lambda_1 - \lambda_2 - \lambda_3
-\lambda_4$ is positive or zero, and (ii) those for which the
same combination is negative.  Again, the numbers $\lambda_i$
are the eigenvalues of the matrix
$R = \sqrt{\sqrt{\rho}\tilde{\rho}\sqrt{\rho}}$.  We consider
class (i) first.

For any density matrix $\rho$ in this class, we will define
successively three specific decompositions of $\rho$, the
last of which will be the optimal decomposition that we seek.
Each of these decompositions consists of exactly $n$ pure states,
$n$ being the rank of $\rho$ as above.  For the system
we are considering,
$n$ is always less than or equal to 4.

The first decomposition consists of states
$|x_i\rangle$, $i=1,\ldots ,n$, satisfying

\begin{equation}
\langle x_i|\tilde{x}_j\rangle = \lambda_i\delta_{ij}.   \label{x}
\end{equation}
The states $|x_i\rangle $ can be said to be
``tilde-orthogonal.'' We obtain such
a decomposition as follows.  First note that if the set
$\{ |x_i\rangle \}$ is defined via an $n \times n$ unitary
matrix $U$ as in Eq.~(\ref{w}), then the ``tilde inner
products'' $\langle x_i|\tilde{x}_j\rangle $ can be
written as

\begin{equation}
\langle x_i|\tilde{x}_j\rangle = (U \tau U^T)_{ij}, \label{Ux}
\end{equation}
where $\tau_{ij} \equiv \langle v_i|\tilde{v}_j \rangle$
is a symmetric but not necessarily Hermitian matrix.
(The states $|v_i\rangle $ are the eigenstates of $\rho$ defined
earlier.)
In order that condition (\ref{x}) be satisfied, we want $U \tau U^T$
to be diagonal.  It happens that for any symmetric matrix $\tau$,
one can always find a unitary $U$ that diagonalizes $\tau$ in this
way \cite{HJ}.  Moreover, the diagonal elements of $U \tau U^T$ can
always be made real and non-negative, in which case they are the
square roots of the eigenvalues of $\tau \tau^{*}$.  (To see how
this works, note that $U$ must diagonalize $\tau \tau^{*}$
in the usual sense; that is, $U \tau \tau^{*} U^{\dag}$ is diagonal.)
The square roots of the eigenvalues of $\tau \tau^{*}$ are the same
as the eigenvalues of $R$, so that
condition (\ref{x}) is fulfilled as long as the diagonalizing
matrix $U$ is chosen in such a way that the numbers $\lambda_i$
appear in their proper order.  Thus one can always find a decomposition
with the desired property.  It is interesting to note that
the vectors $|x_i\rangle$ of this decomposition are
right-eigenvectors of the non-Hermitian matrix $\rho \tilde{\rho}$.
One can see this by writing $\rho$ as $\sum_i |x_i\rangle\langle x_i|$
and using Eq.~(\ref{x}).
We could in fact have used this property to give
an alternative specification of the ensemble $\{|x_i\rangle \}$.

Our second decomposition of $\rho$, which we label $\{ |y_i\rangle \}$,
$i=1,\ldots ,n$,  is hardly different from the first:

\begin{equation}{\normalbaselineskip=20pt\matrix{
&|y_1\rangle = |x_1\rangle ;  \hfill \cr
&|y_j\rangle = i|x_j\rangle {\hskip 1pc}\hbox{for \,} j \ne 1.}}
\label{y}
\end{equation}
It is indeed physically equivalent to the first decomposition,
but the phase factors will become important shortly when
we take linear combinations of these vectors.

The decomposition $\{ |y_i\rangle \}$ typically does not have a
small average entanglement, but it does have a
property that will make it useful for finding an optimal
ensemble.  In order to express this property, let us define
the ``preconcurrence''
$c$ of a pure state $|\psi \rangle$ to be

\begin{equation}
c(\psi) = \frac{\langle \psi | \tilde{\psi} \rangle}{\langle
          \psi |\psi \rangle},
\end{equation}
where we have allowed for the possibility that $|\psi \rangle$ may be
subnormalized.  Note that the preconcurrence is the same as the
concurrence of Eq.~(\ref{C(psi)}) but without the absolute
value sign.  The decomposition $\{ |y_i\rangle \}$ is special in
that its average preconcurrence has the value $C(\rho)$ of
Eq.~(\ref{C(rho)}).  To see this,
recall that the probability of the state $|y_i\rangle$ in the
decomposition is $\langle y_i|y_i \rangle$, so that the average
preconcurrence is

\begin{equation}
\langle c \rangle = \sum_i \langle y_i|y_i \rangle
       \frac{\langle y_i|\tilde{y}_i \rangle}{\langle y_i|y_i \rangle}
       = \sum_i \langle y_i|\tilde{y}_i \rangle.  \label{avec}
\end{equation}
The sum can be evaluated immediately from Eqs.~(\ref{x}) and
(\ref{y}), yielding $\langle c \rangle = \lambda_1 - \lambda_2
- \lambda_3 - \lambda_4 = C(\rho)$.  Here we have used the
fact that if $n < 4$, the numbers
$\lambda_i$ with $i > n$ are all zero.

We would like to find a decomposition that,
like $\{ |y_i\rangle \}$, has $\langle c \rangle
= C(\rho)$, but which also has the property that
the preconcurrence (and hence the concurrence)
of each individual {\em state}
is equal to $C(\rho)$.  It would then follow
immediately that the average entanglement is ${\cal E}(C(\rho))$,
since this would be the entanglement of each state in
the decomposition.
In seeking such a decomposition, we will confine
ourselves to transformations that leave the average
preconcurrence unchanged, and use these transformations to
adjust the individual
preconcurrences until they are all the same.  The result will
be our final decomposition of $\rho$.

Now, any
decomposition with $n$ elements can be written
in terms of the states $|y_i\rangle$ via the equation

\begin{equation}
| z_i \rangle = \sum_{j=1}^n V_{ij}^* | y_j\rangle,  \label{wprime}
\end{equation}
where $V$ is an $n \times n$ unitary matrix.  The average
preconcurrence of the ensemble $\{ |z_i\rangle \}$ is

\begin{equation}
\langle c\rangle = \sum_i \langle z_i|\tilde{z}_i \rangle =
     \sum_i (VYV^T)_{ii} =
     {\mbox{Tr}\,}{(VYV^T)},  \label{avepre}
\end{equation}
where $Y$ is the real diagonal matrix defined by
$Y_{ij} = \langle y_i|\tilde{y}_j \rangle$.  Thus
the average preconcurrence is unchanged by
any {\em real} unitary matrix $V$ (that is, any
orthogonal matrix), since in that case $V^T = V^{-1}$ and the
trace in Eq.~(\ref{avepre}) is preserved.

Even restricting ourselves to orthogonal matrices, we
retain more than enough freedom to make the preconcurrences
of the individual states equal.  One way to do this is
as follows.  First, select the two states $|y_i\rangle$
with the largest and smallest values of the preconcurrence.
Since the set $\{ |y_i\rangle \}$ has the correct {\em average}
preconcurrence, either all the preconcurrences are already
equal to $C(\rho)$, or else the largest one
is too large and the smallest one
is too small (typically negative).  In the latter case,
consider the set of positive-determinant
orthogonal transformations that act only on these two
extreme states as in Eq.~(\ref{wprime}),
changing them into new states that
we call $|z_a\rangle$ and $|z_b\rangle$.  (This set
of transformations is simply the
one-parameter set of rotations in
two dimensions.  It is worth emphasizing, however, that we
are not using them to rotate the vector space; rather, we
are directly forming new linear combinations of the two
specified states.  The other states $|y_i\rangle$
are not changed.)  Among this set of transformations
is one that simply interchanges the two extreme states
and thus interchanges their preconcurrences.
Therefore, by continuity there must exist an intermediate
transformation that makes the preconcurrence
of $|z_a\rangle$ equal to $C(\rho)$.  Perform this
transformation, thereby fixing one element of the ensemble
to have the correct concurrence.  Next, consider
the remaining $n-1$ states, that is, $|z_b\rangle$ and the
remaining $|y_i\rangle$s, and perform the same operation
on them.  Continuing in this way, one finally arrives at
a set of states all having concurrence equal to $C(\rho)$.
This we take to be our final decomposition $\{ |z_i\rangle \}$,
which, as we have argued above, achieves the claimed minimum
average entanglement ${\cal E}(C(\rho))$.
Thus the value of entanglement given in our
formula (\ref{E(rho)}) can always be attained, at least for the
case in which
$\lambda_1 - \lambda_2 - \lambda_3 - \lambda_4 \ge 0$.

We now wish to show that no decomposition of $\rho$ has
a {\em smaller}
average entanglement.  For this it is enough to show that
no decomposition has a smaller average {\em concurrence}:
the average entanglement cannot be less than
${\cal E}(\langle C \rangle)$ because of the convexity of the
function ${\cal E}$.  Now, the average concurrence
of a general decomposition is given by an equation similar
to Eq.~(\ref{avepre}) but with an absolute value sign:

\begin{equation}
\langle C\rangle = \sum_i |(VYV^T)_{ii}|.  \label{aveC}
\end{equation}
Here $V$ is an $m \times n$ matrix whose $n$ columns are orthonormal
vectors.  The dimension $m$ of these vectors can be arbitrarily
large, since the decomposition may
consist of an arbitrarily large number of pure states (though
prior results guarantee that one need not consider values of
$m$ larger than sixteen \cite{16}).
In terms of the components of
$V$ and $Y$, we can write the average concurrence as

\begin{equation}
\langle C\rangle = \sum_i \Bigl| \sum_j (V_{ij})^2 Y_{jj} \Bigr|.
\end{equation}
To obtain the desired lower bound on this sum, we need use only the
fact that $\sum_i |(V_{ij})^2| = 1$.  That is, we can show that
for any complex numbers $\alpha_{ij}$ such that
$\sum_i |\alpha_{ij}| = 1$, we have

\begin{equation}
\sum_i \Bigl| \sum_j \alpha_{ij} Y_{jj}\Bigr| \ge \lambda_1-\lambda_2
    -\lambda_3 -\lambda_4.
\end{equation}
The proof is straightforward: first note that there is
 no loss of generality
in taking each $\alpha_{i1}$ to be real and positive.
(The phases of the other
$\alpha_{ij}$s can be changed to compensate.)  Then we can say

\begin{equation}{\normalbaselineskip=20pt\matrix{
\sum_i | \sum_j \alpha_{ij} Y_{jj} |
     &\ge | \sum_{ij} \alpha_{ij} Y_{jj} | \hfill \cr
     &=|\lambda_1-\sum_{j=2}^n (\sum_i\alpha_{ij}) \lambda_j|\hfill\cr
     &\ge \lambda_1 - \lambda_2 -\lambda_3 -\lambda_4  \hfill  \cr
     &= C(\rho).  \hfill}}
\end{equation}
(Again we are using the fact that any $\lambda_j$ with $j > n$ is zero.)
Thus no decomposition of $\rho$ can achieve an average concurrence lower
than $C(\rho)$ or an average entanglement lower than ${\cal E}(C(\rho))$.

There remains one case to consider, namely, density matrices for which
$\lambda_1 - \lambda_2 - \lambda_3 - \lambda_4 < 0$.  For these
density matrices
our formula predicts that the entanglement should be zero, that is,
that there should be a decomposition of $\rho$ into unentangled pure
states.  To show that this is indeed the case, we again start with
the decomposition $\{ |x_i\rangle \}, i = 1,\ldots ,n$, of Eq.~(\ref{x}).
If $n$ is equal to 3---the values $n=1$ and $n=2$ are not
possible for the case we
are now considering---it is convenient to supplement this set with
a dummy state
$|x_{4}\rangle$ equal to the
zero vector.  From the complete set we directly form
our final decomposition $\{ |z_i\rangle \}$:

\begin{equation}{\normalbaselineskip=20pt\matrix{
&|z_1\rangle = \frac{1}{2}(e^{i\theta_1}|x_1\rangle +
     e^{i\theta_2}|x_2\rangle + e^{i\theta_3}|x_3\rangle +
     e^{i\theta_4}|x_4\rangle) \hfill \cr
&|z_2\rangle = \frac{1}{2}(e^{i\theta_1}|x_1\rangle +
     e^{i\theta_2}|x_2\rangle - e^{i\theta_3}|x_3\rangle -
     e^{i\theta_4}|x_4\rangle) \hfill \cr
&|z_3\rangle = \frac{1}{2}(e^{i\theta_1}|x_1\rangle -
     e^{i\theta_2}|x_2\rangle + e^{i\theta_3}|x_3\rangle -
     e^{i\theta_4}|x_4\rangle) \hfill \cr
&|z_4\rangle = \frac{1}{2}(e^{i\theta_1}|x_1\rangle -
     e^{i\theta_2}|x_2\rangle - e^{i\theta_3}|x_3\rangle +
     e^{i\theta_4}|x_4\rangle) , \hfill}}
\label{yprime}
\end{equation}
where the phase factors are chosen so that

\begin{equation}
\sum_j e^{2i\theta_j}\lambda_j = 0. \label{null}
\end{equation}
Such phase factors can always be found when
$\lambda_1 < \lambda_2 + \lambda_3 + \lambda_4$ ($\lambda_1$ being
the largest of the four numbers as always).
The condition (\ref{null}) together with the property (\ref{x}) of
the set $\{ |x_i\rangle \}$ guarantee that each state $|z_i\rangle$
has zero concurrence and hence zero entanglement.  This completes
the proof of the formula (\ref{E(rho)}).

Our formula makes possible the easy evaluation of entanglement of
formation for a pair of qubits, and should thus
facilitate the investigation
of any number of questions concerning entanglement.  However, there
remains a very basic question concerning the {\em interpretation}
of the entanglement of formation that has not yet been resolved.
For any pure state $|\psi \rangle $ of a bipartite
system, the entanglement $E(\psi)$ defined
in Eq.~(\ref{pure}) has a very simple and
elegant interpretation \cite{interp}: if two separated observers
Alice and Bob start out with no shared entanglement, then in order
for them to create many pairs in the state $|\psi \rangle $,
such that Alice ends up with
one member of each pair and Bob has the other, it is
necessary that for each pair produced, at least
$E(\psi)$ qubits must pass across
an imaginary plane separating Alice and Bob; moreover, as the number
of pairs approaches infinity, the number of transmitted
qubits needed per pair can be made arbitrarily close to $E(\psi)$.
That is,
$E(\psi)$ measures the amount of quantum information that must be
exchanged between Alice and Bob in order to create the state
$|\psi \rangle $.  It seems likely that one can apply the same
interpretation to the entanglement of formation of a {\em mixed}
state \cite{formation}, but this conclusion
depends on a property of $E(\rho)$ that
has not yet been demonstrated \cite{Popescu}.  The question
is whether $E(\rho)$ is
{\em additive}, that is, whether, if Alice and Bob have
$n$ pairs in the state
$\rho$, the entanglement of formation of that whole system
is exactly $n$ times the entanglement of formation of a single pair
and not less.
In mathematical terms, the issue is whether $E(\rho^{\otimes n})
= nE(\rho)$.  It is conceivable that the formula proved in this
Letter will help to settle this question
in the case of qubits,
but more likely an entirely different and more general argument
will have to be found.  If it is determined that $E(\rho)$ is
indeed additive, then this finding will considerably strengthen the
physical interpretation of our formula.

I would like to thank a number of colleagues whose comments
and suggestions have been of great help in this work:
Valerie Coffman, Scott Hill, Joydip Kundu,
Hideo Mabuchi, Michael Nielsen, David Park,
Eric Rains, John Smolin, Ashish Thapliyal, and especially Chris Fuchs.

\end{document}